\documentstyle[12pt,moriond,graphicx]{article}
\begin{document}

\vspace*{-10mm}
\noindent
To appear in {\em Dwarf Galaxies and Cosmology}, the Proceedings of the 
XVIIIth Moriond Astrophysics Meeting, Les Arcs, Savoie, France, 
14th--21st March, 1998, eds. T.~X.~Thu\^{a}n, C.~Balkowski, V.~Cayatte 
and J.~Tr\^{a}n Thanh V\^{a}n, publ. Editions Fronti\`{e}res, 
Gif-sur-Yvette.\\[-10mm]

\heading{DWARF SPHEROIDAL GALAXIES \\ IN THE VIRGO CLUSTER}

\author{J. B. Jones $^{1}$, S. Phillipps $^{1}$, 
    J. M. Schwartzenberg $^{1}$, Q. A. Parker $^{2}$}
    {$^{1}$ Astrophysics Group,  Department of Physics, 
    University of Bristol,\\ 
    Tyndall Avenue, Bristol, BS8 1TL, United Kingdom.}
    {$^{2}$ Anglo-Australian Observatory, Siding Spring, Coonabarabran,\\ 
    New South Wales 2357, Australia.}

\begin{moriondabstract}
We present a study of the smallest and faintest galaxies found in 
a very deep photographic R band survey of two regions of the Virgo 
Cluster, totalling 3.2 square degrees, made with the UK Schmidt 
Telescope. The objects we detect have the same physical sizes and 
surface brightnesses as Local Group dwarf spheroidal galaxies.
The luminosity function of these extremely low luminosity galaxies (down
to ${\rm M}_{\rm R} \simeq -11$ or about $5 \times 10^{-5} L_{\star}$) 
is very steep, with a power law slope $\alpha = -2.2$.
\end{moriondabstract}

\section{Introduction}

    The properties of the faint end of the luminosity function 
are poorly constrained outside the Local Group, both in terms 
of the numbers and characteristics of the galaxies. Some studies 
have found evidence for very steep luminosity functions in both 
cluster~\cite{sdp97}\cite{t98} and field~\cite{lvdy97} environments. 
The importance of very low luminosity galaxies in understanding 
galaxy formation and evolution~\cite{t98} demands that progress 
is made in identifying and studying low luminosity dwarfs in new 
environments~\cite{cadcs98}. 

    This survey extends the study of extremely low luminosity 
galaxies to the Virgo Cluster (to M$_{\rm R} = -11$~mag), into 
the regime of Local Group dwarf spheroidals. 
It extends the definitive study of the Virgo 
Cluster by Binggeli, Sandage~\& Tammann~\cite{bst85} by 3~mag 
beyond their completeness limits and complements 
the survey of large, low surface brightness galaxies of Impey, Bothun~\& 
Malin~\cite{ibm88}. By coadding multiple photographic exposures 
with the UK Schmidt Telescope, galaxies are detected having 
central surface brightnesses as faint as 25~R~mag~arcsec$^{-2}$, 
equivalent to 26.5~B~mag~arcsec$^{-2}$. 
We report here results from an initial survey~\cite{ppsj98} 
covering~3.2~deg$^2$.

\section{The Data and Image Detection}

    The data used here~\cite{ppsj98}\cite{sch96} are part of a larger 
photographic survey~\cite{spp95} of a ten-degree square region of the 
Virgo Cluster using the extremely fine 
grained, highly efficient Tech Pan film on the 1.2m UK Schmidt 
Telescope~\cite{pp93}. Six individual long (1 to 1.5 hour) exposures 
of the same area, the South-East quadrant of the Virgo Cluster, 
were scanned with the SuperCOSMOS automatic measuring
machine at the Royal Observatory Edinburgh. Two 
$1.3^{\circ} \times 1.3^{\circ}$ subregions were used for the 
initial survey, one near the cluster centre, one further out. 
The digitised data from the six separate films were sky subtracted 
and were matched in intensity~\cite{spp96} 
before coaddition using median stacking. This provided an equivalent 
exposure time of about 7 hours, while the coadded data are free of 
artefacts such as emulsion 
defects, dust particle images and satellite trails that affect only 
one film. 

    Image detection was performed with a connected-pixel algorithm 
using a detection threshold $\mu_{lim} = 25.5$ mag arcsec$^{-2}$ 
and a minimum area above the threshold of $A_{lim} = 11$ pixels. 
Each image thus had a minimum S/N ratio of 10 and has 
a magnitude $\mbox{R} \leq 22$. This detection limit is 
0.3~mag~arcsec$^{-2}$ fainter than previous surveys. 
Around 28~000 images were detected in both fields and the large minimum area
ensured that few were spurious. This was confirmed by comparison of a
small subregion of the photographic field with a CCD image taken on the
Anglo-Australian Telescope: 160 of 162 images visible on the film
were matched on the CCD frame, including all those which contributed 
to determining the luminosity function in Section~4. Scale lengths 
and surface brightnesses have been measured for each image 
from the isophotal signals and areas at the detection threshold, assuming 
exponential light profiles.

\section{Identifying Virgo dwarfs}

    Virgo Cluster dwarf elliptical and dwarf spheroidal galaxies can be 
characterised by their image sizes and surface brightnesses: they are 
larger in general than the majority of background images at a given 
central surface brightness. As illustrated in Figure~1, it is possible 
to use the position of images in the magnitude~-- surface brightness 
plane to identify a sample rich in Virgo dwarfs. Figure~1a compares the 
distribution of images in a Virgo field with that of the background 
population. 
The background data are based on AAT CCD observations of a field 
around the South Galactic Pole. The background population dominates 
strongly at scale lengths $\alpha \lsim 1.8$~arcsec. In contrast, 
a majority of galaxies having scale lengths $\alpha > 2$~arcsec 
and central surface brightnesses $\mu_0 > 22.0$~mag~arcsec$^{-2}$ 
are expected to be Virgo dwarfs.

\begin{figure}[htb]
\vspace{-5mm}
\hspace*{-5mm}  \includegraphics[width=97mm]{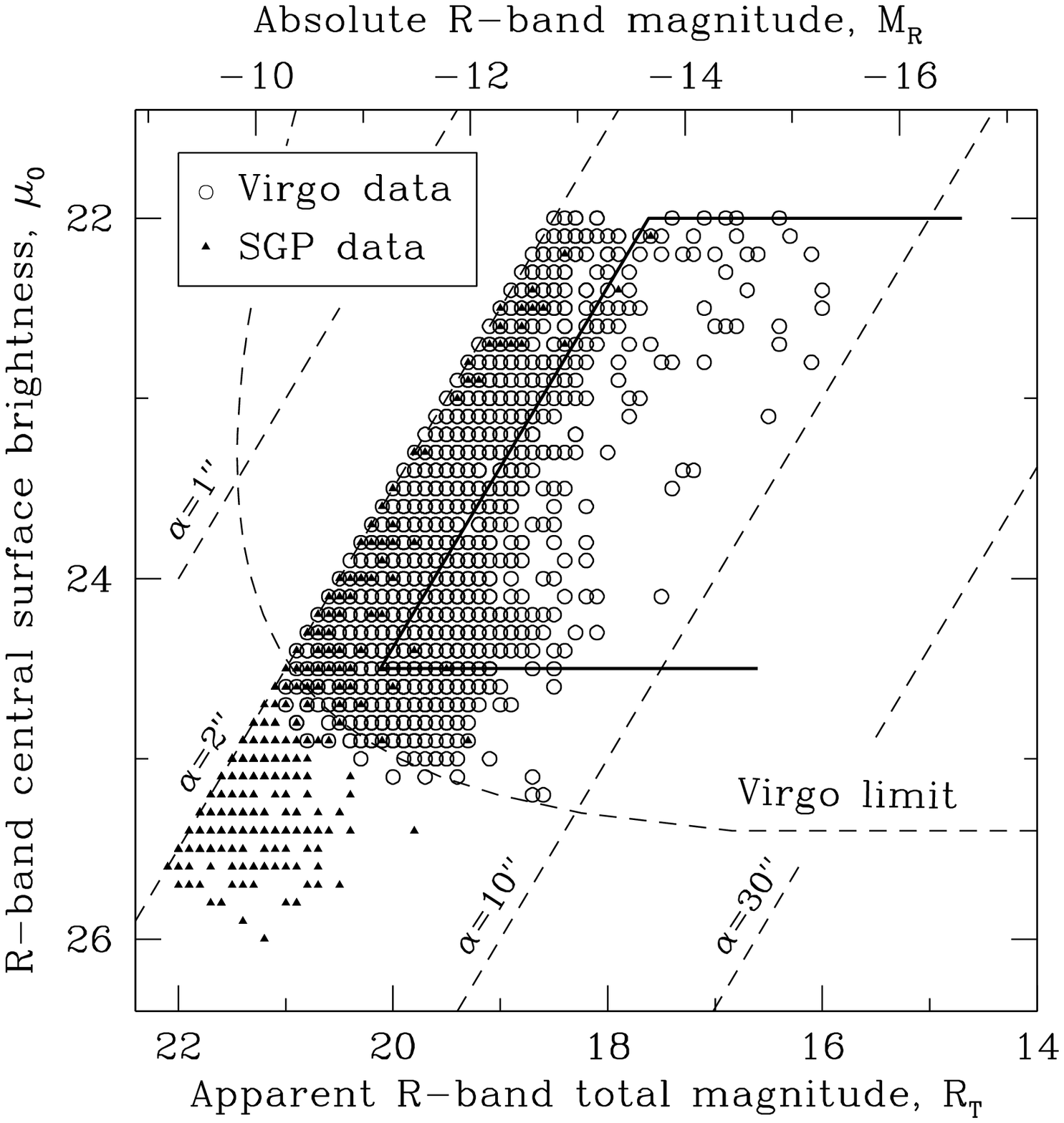} 
\hspace*{-22mm} \includegraphics[width=97mm]{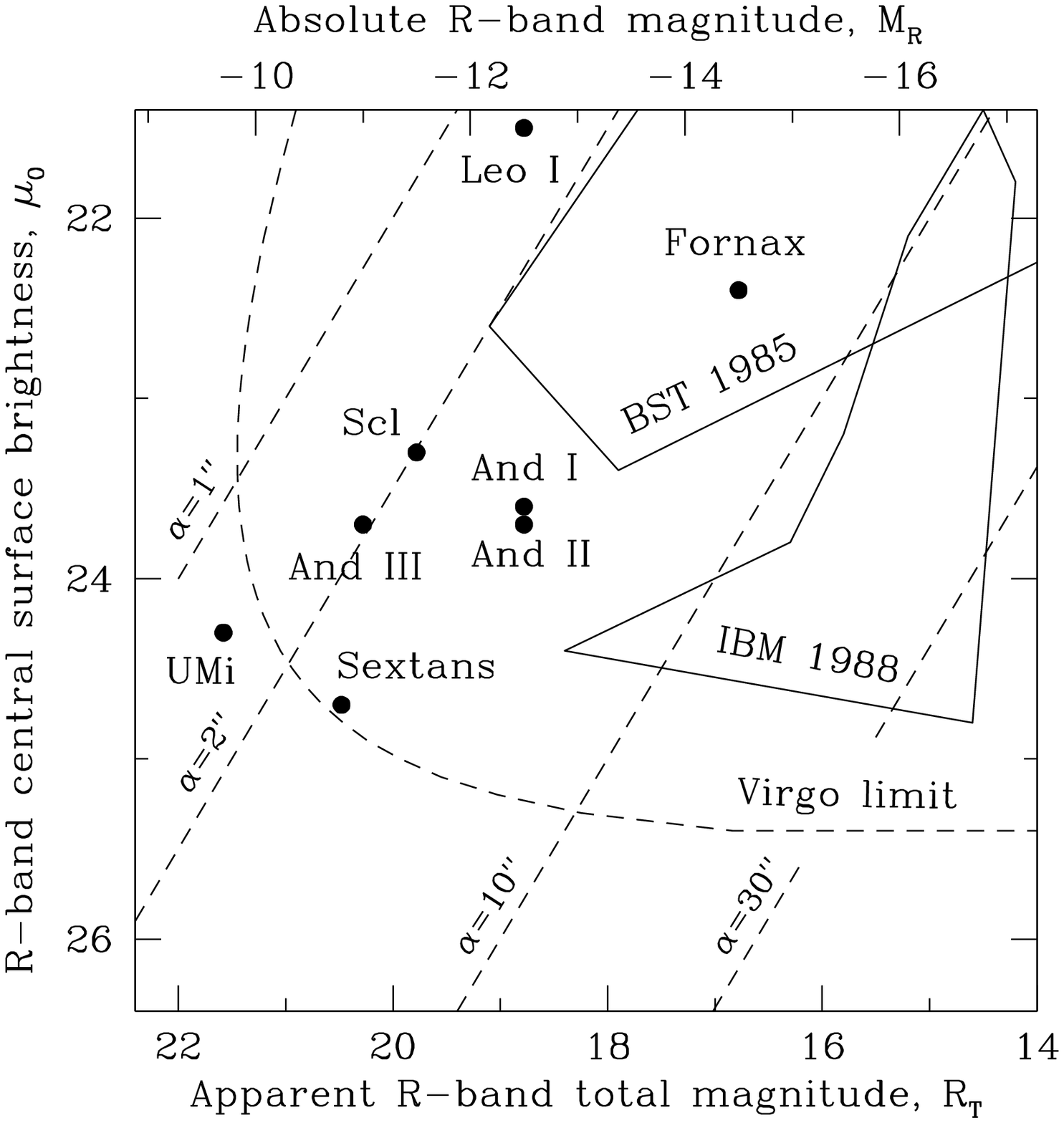} \\[-13mm]
\hspace*{0mm} {\bf (a)} \hspace*{79mm} {\bf (b)} \\[6mm]
\renewcommand{\baselinestretch}{1.0}
{\bf Figure~1: (a)} The magnitude~-- central surface brightness plane 
for galaxies detected in the Virgo Schmidt survey compared with the 
SGP CCD field used for background subtraction. For clarity, galaxies 
having scale lengths $\alpha < 2$ arcsec (where the background 
population dominates) have not been plotted. 
Loci are shown corresponding to exponential galaxies having 
scale lengths $\alpha$ = 1, 2, 10 and 30 arcsec. 
The selection boundary 
imposed on the Schmidt data by the detection methods is shown  
as the dashed curve. The bold lines show the limits of the sample 
used to determine the luminosity function in Section~4.  
{\bf (b)} As Figure~1a, but showing the properties Local Group dSphs 
would have if seen at the distance of the Virgo Cluster. The solid 
lines mark the regions of the plane covered by the Binggeli 
{\it et al.}~\cite{bst85} and Impey {\it et al.}~\cite{ibm88} samples. 
Throughout this paper we adopt a distance modulus of 31.3~mag for the 
cluster.\\
\renewcommand{\baselinestretch}{1.5}
\end{figure}

    The properties of the observed galaxies can be compared with those 
of Local Group objects. Figure~1b shows the positions that Local Group 
dSphs~\cite{casdc92} would occupy in the magnitude~-- surface 
brightness diagram were they located at the distance of the 
Virgo Cluster: many of the local dSphs could be distinguished 
unambiguously from the background population. For comparison, 
the regions occupied by the galaxies in the samples of 
Binggeli, Sandage~\& Tammann~\cite{bst85} and of Impey, Bothun~\& 
Malin~\cite{ibm88} are shown in the figure; neither of these other 
studies reaches the small, low surface brightness galaxies found here.

\section{The luminosity Function}

    Information about number densities of genuine Virgo Cluster 
galaxies of various types can be derived by subtracting the background 
contribution from the total number of observed galaxies at each point 
in the magnitude~-- surface brightness plane. This technique has been 
used to determine the luminosity function for a subsample of Virgo dwarfs, 
defined by $\alpha \geq 3.0$~arcsec and central surface brightnesses 
$22.0$~mag~arcsec$^{-2} \leq \mu_0 \leq 24.5$~mag~arcsec$^{-2}$, 
in each of the two fields studied. This scale length limit ensures 
that the background galaxy contamination is minimal (about 4\%),
reducing difficulties which might be caused by comparing data 
having different point-spread functions. These sample limits are 
shown in Figure~1a. 

    The luminosity functions~\cite{ppsj98} for the two fields are 
presented in 
Figure~2. The inner field is centred close to M87 and samples the 
population in the core of the cluster. The area immediately around 
M87, having a higher effective background light level, has been 
excluded from the analysis; the area surveyed is 1.58~deg$^2$. The 
outer cluster field lies $3.1$~deg to the south-southeast of M87 and 
is 1.61~deg$^2$ in area. The inner and outer cluster fields have 
very similar 
luminosity functions. Both have steep faint-end slopes: the combined 
data set has $\alpha = -2.26 \pm 0.13$ over the  magnitude range 
$15.5 < \mbox{R} < 20.0$. The downturn at the last low luminosity 
point is a consequence of the sample limits (defined by scale length 
and surface brightness, not by magnitude). 
At the bright end, the luminosity functions 
are consistent with that of Binggeli, Sandage~\& 
Tammann~\cite{sbt85}\cite{bst88} (although a comparison is 
complicated by the slightly different samples and filters used 
in the two surveys). The density of galaxies is marginally 
greater in the outer cluster field than in the core (by a factor of 
30\%), which is consistent with a depletion of dwarfs in the 
core, as has been observed in the Coma Cluster~\cite{tg93}. 

     The survey is currently being extended to the full ten-degree 
square region centred on the cluster core. 
Photographic data in an off-cluster field will provide improved background 
subtraction, allowing a luminosity function to be constructed which 
extends further into the region of Figure~1 occupied by field galaxies 
than has been possible here.

\begin{figure}[thb]
\vspace{4mm}
\hspace*{45mm} M$_{\rm R}$ \hspace*{77mm} M$_{\rm R}$ \\[2mm]
\hspace*{2mm} \includegraphics[width=77mm]{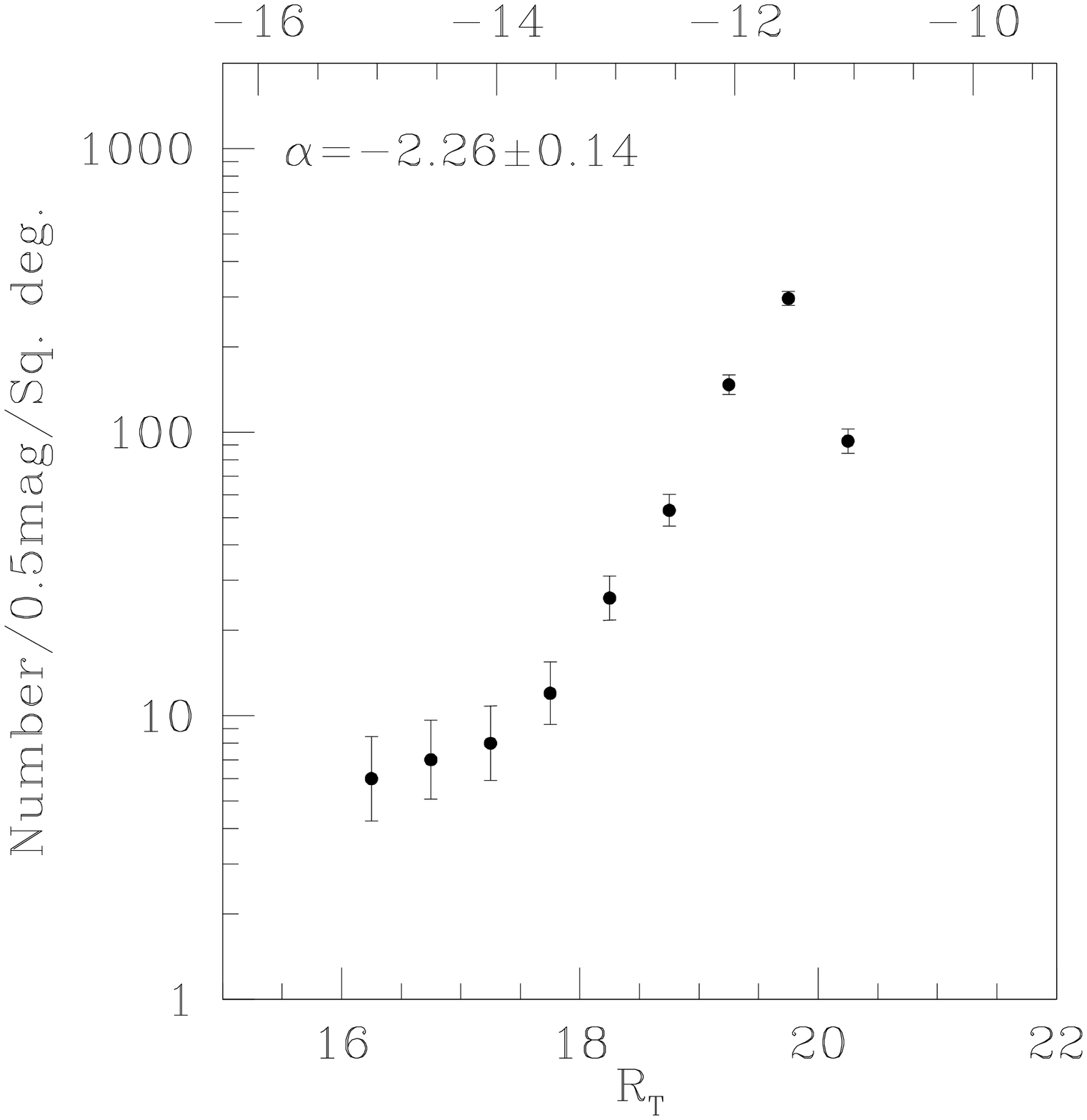} 
\hspace*{8mm} \includegraphics[width=77mm]{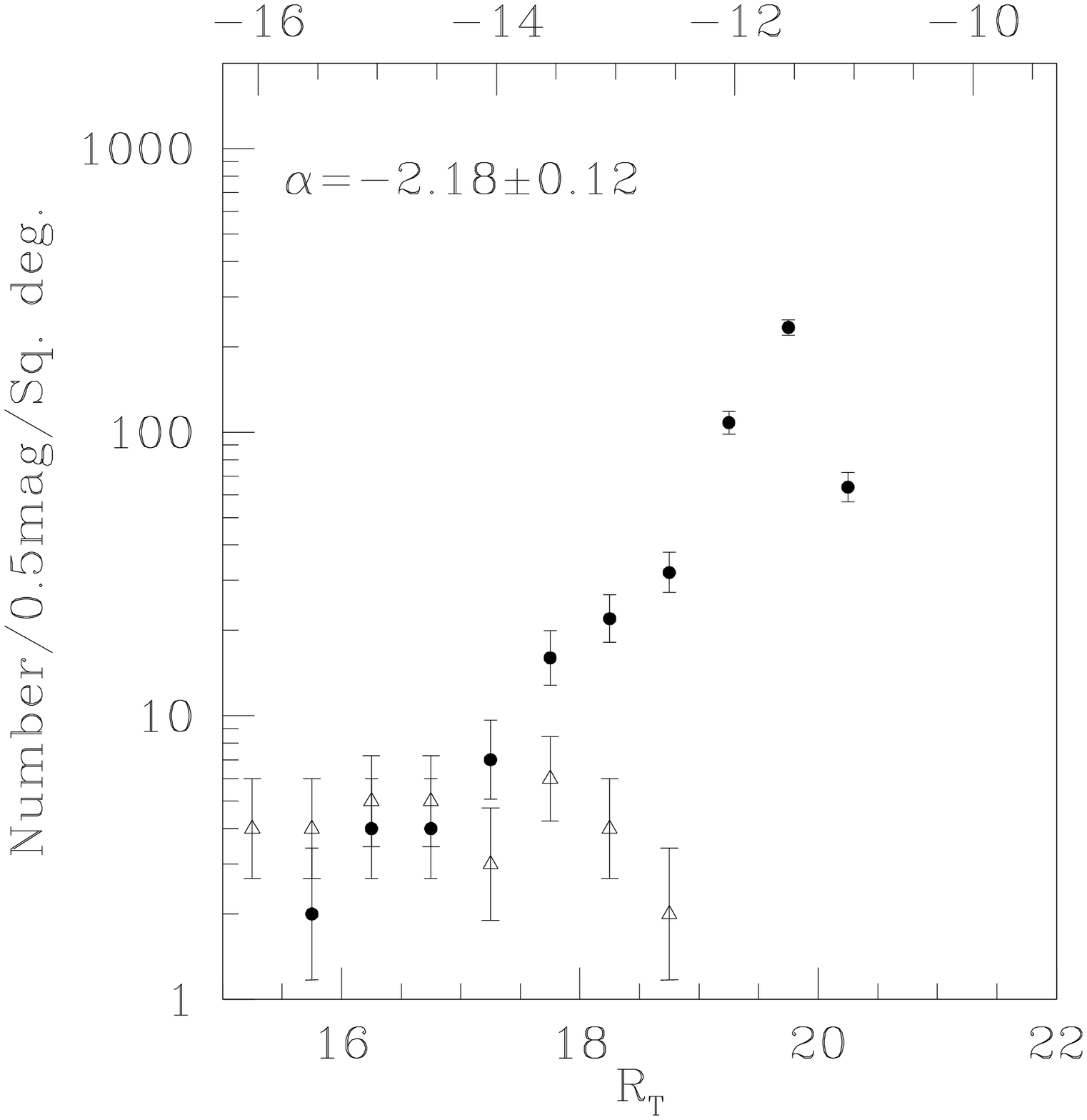} \\[-7mm]
\hspace*{3mm} {\bf (a)} \hspace*{79mm} {\bf (b)} \\[4mm]
\renewcommand{\baselinestretch}{1.0}
{\bf Figure~2:} The absolute magnitude distribution of the sample 
of Virgo dwarfs of Section~4 for the (a) outer and (b) inner fields. 
Error bars shown are based on Poissonian statistics. The luminosity 
function of Sandage, Binggeli \& Tammann~\cite{sbt85}\cite{bst88} 
is also shown (open triangles) for the Virgo Cluster Catalog galaxies 
which overlap the inner cluster field (assuming 
$\mbox{B}-\mbox{R} = 1.5$). 
\renewcommand{\baselinestretch}{1.5}
\end{figure}

\vfill
\end{document}